\documentclass[submission, Phys]{SciPost}

\begin{document}

\begin{center}{\Large \textbf{
Vision for QCD to the 2030s and Beyond\footnote{Based on the closing talk at the 28th Workshop
on Deep-Inelastic Scattering and Related Subjects, Virtual Event, Stony Brook, April 12-16, 2021.}
}}
\end{center}

\begin{center}
George  Sterman 
\end{center}

\begin{center}
C.N.\ Yang Institute for Theoretical Physics and Department of Physics and Astronomy\\
Stony Brook University, Stony Brook New York 11794-3840, U.S.A.
george.sterman@stonybrook.edu
\end{center}

\begin{center}
\today
\end{center}


\section*{Abstract}
{\bf
In this closing talk of the DIS 2021 Workshop, I review some of the lessons we've learned about quantum chromodynamics,
and reflect on what we may hope to learn in the coming years.
}

\vspace{10pt}
\noindent\rule{\textwidth}{1pt}
\tableofcontents\thispagestyle{fancy}
\noindent\rule{\textwidth}{1pt}
\vspace{10pt}

\section{Introduction}
\label{sec:intro}
I have been asked to provide a vision in this talk, but it isn't meant to be an exercise in prediction.  The year 2030
     is still a long way off.  The study of quantum chromodynamics, 
     however, is now driven as much by evolving experimental capabilities as by the evolution of theory,
     so we have some sense of what may happen in the meantime.

Ten years ago, the last of the great Standard Model discovery machines, 
    the Tevatron, closed up shop, following LEP and HERA, while
    RHIC entered its second fruitful decade.  The decade just past also 
    saw the historic LHC Runs I and II, as CEBAF transitioned 
    from 6 to 12 GeV at Jefferson Lab.   All these milestones were reflected in the DIS series of conferences.

Starting with RHIC, many accelerator capabilities
    have been designed with QCD in mind, at JLab of course, and in
    the decade unfolding, the EIC.   The LHC was not built
    for QCD, but the insightful designs of its detectors make it
    (of necessity) a powerful QCD discovery tool.
    
    Over the past twenty years, QCD has brought nuclear and
    particle physics (back) together.  Roads from Newport News
    and Upton lead to Geneva (and back).
The specifically QCD experimental capabilities that will link
    the 2020s and the 2030s, including fixed target experiments at
    Fermilab, JLab, CERN and Brookhaven, have already
    paved the way for the nascent Electron Ion Collider project, based
    on the demonstrated need for high statistics to reveal
    the structure of the nucleon, and high energy to unlock
    the dense gluonic matter from which the mass of the visible
    universe is primarily generated.
Those same energy and statistics will provide the means to study
    the emergence of hadronic from partonic matter.  

The story that follows is one sketch of some lessons we have learned
    about the role QCD in nature, and some we may hope to learn in the
    coming years.   Most of the topics I'll touch on can be found,
    with many references, in the survey \cite{Accardi:2012qut} and recent dedicated study  \cite{AbdulKhalek:2021gbh}
    of QCD physics at the Electron-Ion Collider.   The limited selection of references provided below reflects my own recent reading and knowledge of work that illustrates the discussion that follows.   It is by no means meant to review all the important work that has been done in these areas.

\section{QCD in the Grand Scheme of Things}
\label{sec:grand scheme}

Very early in the history of the universe we know, quarks and gluons secluded themselves to a nearly vanishingly small, and ever-decreasing, proportion of space, occupying something like
one $10^{-45}${\it th}  of the volume of the observable universe.   The reason, so far as we understand it, 
lies in the nonabelian phase invariance of the
quarks:  three utterly indistinguishable
colors, connected by gluonic excitations.
This, of course, is quantum chromodynamics. 

From inside the nucleon, quarks send and receive signals to and from the outside world, using the rest of the Standard Model.  Gluons are very
busy, of course, but they let the quarks and antiquarks do all the talking.   Nucleons and nuclei give electrons a reason 
to stick around and form the world we can see.  But the QCD degrees of freedom are always
available, lurking in the vacuum, ready to lend a hand and work alongside the rest of the Standard Model, whenever enough energy arrives in the neighborhood.

\begin{figure}[h]
\centering
 \includegraphics[scale=0.5]{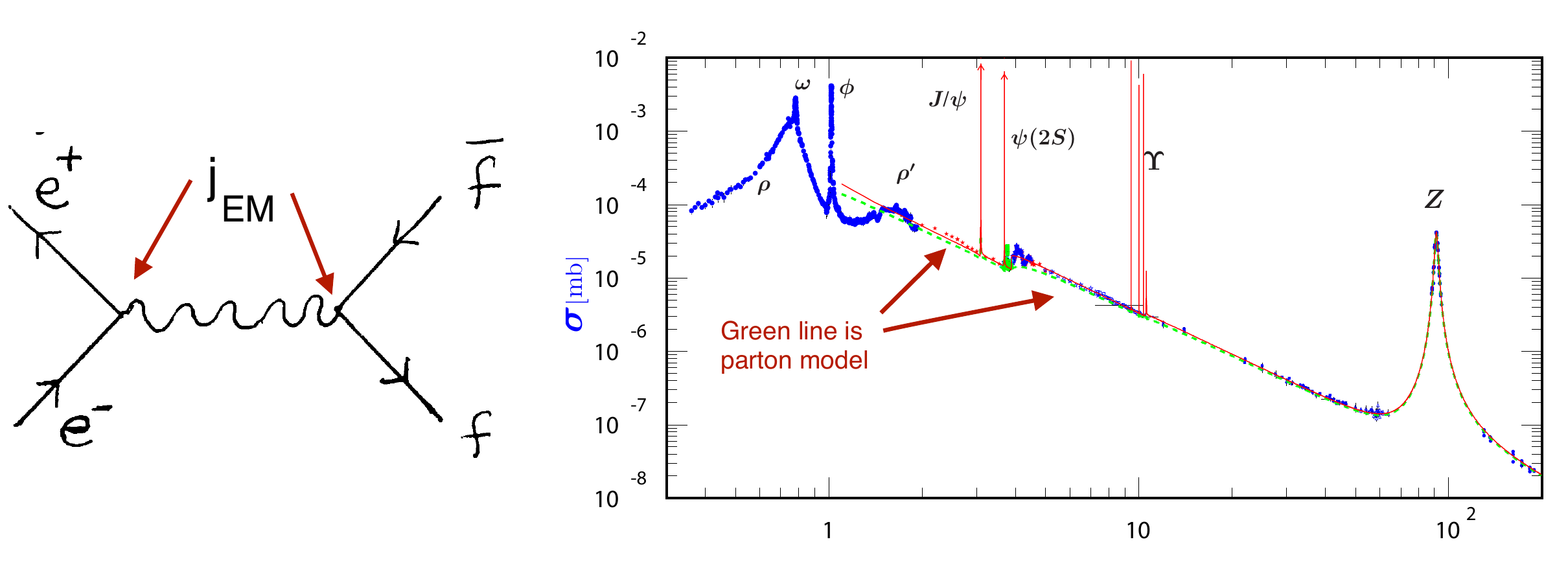} 
 \caption{Left:  The electroweak process mediating the annihilation of lepton pairs to hadrons.  Right: The total cross section for electron-positron annihilation to hadrons \cite{ParticleDataGroup:2020ssz}.}
\label{fig:epem}
\end{figure}



The very same theory possesses asymptotic 
freedom, which has opened windows to its
fundamental degrees of freedom.  Just knowing that QCD is asymptotically free 
is enough to get a good estimate of an important
physical process, the total $\rm e^+e^-$ annihilation cross section to hadrons, as if the theory were free.   The familiar electroweak
process mediating the transition of electron-positron pairs to quark pairs, and hence to the world of hadrons, is shown on the left of Fig.\ \ref{fig:epem}.
The inclusive cross section, on the right of Fig.\ \ref{fig:epem}, is very much the same as what it would have been if the quarks were able
to go their merry ways, without further interactions.   
At long distances, the same theory forms only bound states, the mesons and baryons of the quark model, and, as discovered
in recent years, exotics of a surprisingly wide variety \cite{Karliner:2017qhf}.    A few of the quark model states are visible on the right of the figure.
Yet all of these ``long-distance" phenomena do not radically change the {\it inclusive} annihilation cross section.

\section{Looking Closely, and from Afar}
\label{sec:closely}

Much of our knowledge of QCD comes from a judicious choice of quantities to observe.   Inclusive cross sections
offer a window into short-distance processes, while providing a partonic map of the structure of hadrons.    
Inclusive, of course, is a relative term, and the observation and analysis of the more fine-grained structure of final states,
particularly involving jets of hadrons, affords a variable resolution into the long-time behavior of the theory.   This is the case perturbatively, and this approach
may some day provide a theory of the transition from partonic to hadronic descriptions of strongly-interacting matter.

\subsection{Inclusive and exclusive observables and factorization}

In a broad sense, the ``universal" form of hard-scattering observables, both cross sections and amplitudes, is a
{\it factorization}, very schematically of the form   \cite{Collins:1989gx,Sterman:2014nua}
\begin{equation}
S\ =\ C\ \times \ F\, .
\label{eq:fact-scheme}
\end{equation}
The left-hand side is an observable quantity, while the right-hand side separates a calculable
``short distance" factor, represented by $C$, which can be computed in perturbation theory, from a ``long distance" quantity, $F$ accessible
only to experimental determination, and increasingly, to numerical simulation.   The product represents in general a convolution,
most simply in partonic momentum fraction.

\begin{figure}[h]
\centering
\includegraphics[scale=0.30]{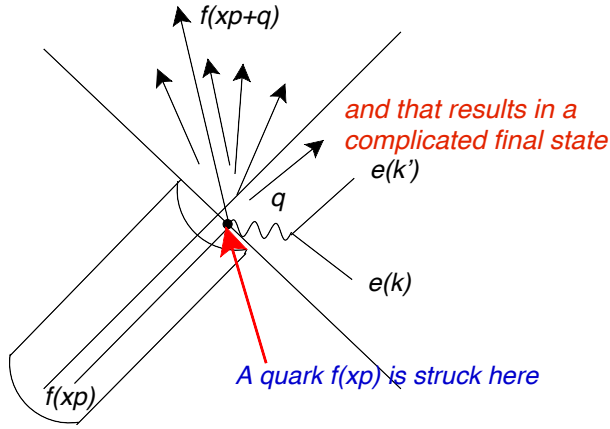}
\caption{A typical event in deep-inelastic scattering.   }
\label{fig:dis}
\end{figure}

\begin{figure}[h]
\centering
\includegraphics[scale=0.50]{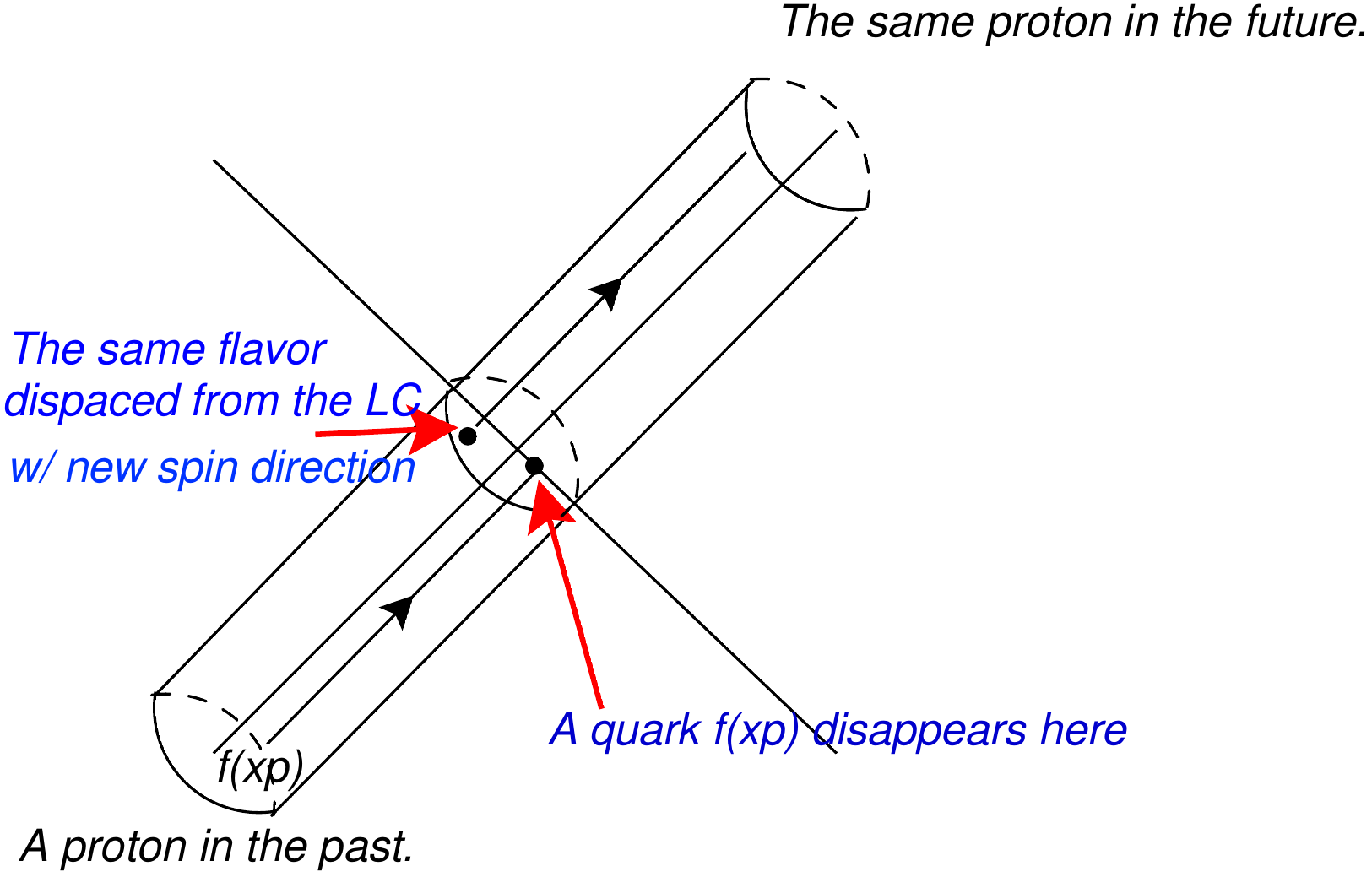}
\caption{Illustration of the matrix element for a collinear or transverse momentuum distribution.   A quark is removed from the proton, and reinserted at a point on or near the
opposite-moving light cone.}
\label{fig:tmd}
\end{figure}

The classic example of Eq.\ (\ref{eq:fact-scheme}) is, of course, deep-inelastic scattering, illustrated by Figs.\ \ref{fig:dis} and \ref{fig:tmd}.   
Figure \ref{fig:dis} illustrates a typical DIS event.  In the figure, the diagonal axes represent the light cone, around which the extended proton, of momentum $p$ and
represented as a cylinder, arrives.  A quark of flavor $f$ and fractional momentum $x$ is suddenly scattered by the exchange of a photon of
momentum $q$.   The full final state is generally complex, and involves many other particles emerging from the fractured proton.   

Despite this complexity,
we can measure the momentum transfer simply by observing the final-state electron.  From its final-state momentum, $k'$ in the figure, we can determine the momentum
transfer $q$.   A large, spacelike momentum transfer $q$ localizes the hard scattering to points that differ in the amplitude and the complex conjugate
amplitude by a nearly lightlike distance.  This provides a clean separation between initial and final states.    In fact, we can do even better by a 
judicious observation of a particle or jet in the direction of the scattered quark's momentum, $p'=xp+q$ in the figure.   In this case, the difference
betweent the scattering event in the amplitude and complex conjugate remains near the lightcone, but deviates slightly in the transverse plane. 
Appropriate sums over 
final states provide
measurements of
parton correlations in the nucleon, including spin.

In all of these cases, fully or semi inclusive, a sum combines the total probabilities of specific final states.   The total result must add up to unity, even if we cannot calculate the probability for any final state individually.  
Corrections arise only from distances of order $1/\sqrt{q^2}$, and can hence be computed perturbatively,  Long-distance contributions
from final states, which are infrared divergent in perturbation theory, cancel.   The result is the ``coefficient" function $C$ in Eq.\ (\ref{eq:fact-scheme}).

The amplitudes for the noncancelling initial states can be interpreted as a parton distribution \cite{Kovarik:2019xvh}, illustrated in Fig.\ \ref{fig:tmd}.   Here, a proton comes in from the past to some fixed point in
spacetime, where a quark is removed from it.   A quark of the same flavor appears at a nearly (but in general not quite) lightlike separation.   The parton distribution is
the amplitude for this proton to reemerge and travel undisturbed into the future.   As suggested in Fig.\ \ref{fig:tmd}, if the spin carried by the quark that disappears
is correlated with the spin of the one that appears, we have a polarization-dependent distribution.    All of these parton distributions can be written as field-theoretic matrix
elements.   For fully inclusive DIS, we need the``collinear" parton distributions 
\begin{equation}
F_Q(x)\ =\ \int \frac{d\lambda}{2\pi}\, e^{-i\lambda\, xP\cdot n} \langle  P |\, \bar{Q} (\lambda n)\, \Gamma\, Q(0)\,  | P \rangle \, ,
\label{eq:pdf}
\end{equation}
where $Q(x)$ represents a quark (or gluon) field, and
where the vector $n^\mu$ is the light cone direction just opposite to the proton in the center of mass.   Here $\Gamma$ represents a Dirac or vector projecction and,
for the experts, an  appropriate gauge link.   When transverse momenta are observed in the final state, we translate the field $\bar{Q}$ off the light
cone as in the figure, and Fourier transform in that ``impact parameter" distance, giving a transverse-momentum dependent, or ``TMD" distribution.   The factorization
of Eq.\ (\ref{eq:fact-scheme}) is then in transverse as well as longitudinal, degrees of freedom \cite{Aybat:2011zv,Collins:2016hqq}.

Factorizations like Eq.\ (\ref{eq:fact-scheme}) with parton distributions of the type in Eq.\ (\ref{eq:pdf}) are  only the first term in a series
 expansion in momentum transfer.   So-called ``higher twist"
terms involve more fields, and depend on matrix elements  such as \cite{Balitsky:2017flc,Koike:2019zxc}
$ \langle  P | \bar{Q} (x) \Gamma G(y) Q(0) | P \rangle$,
with $Q$ a quark and $G$ a gluon field.
For some observables, especially involving spin, this is the leading effect, offering information on short-distance correlations between partons of different flavor.

\begin{figure}[h]
\centering
\includegraphics[scale=0.40]{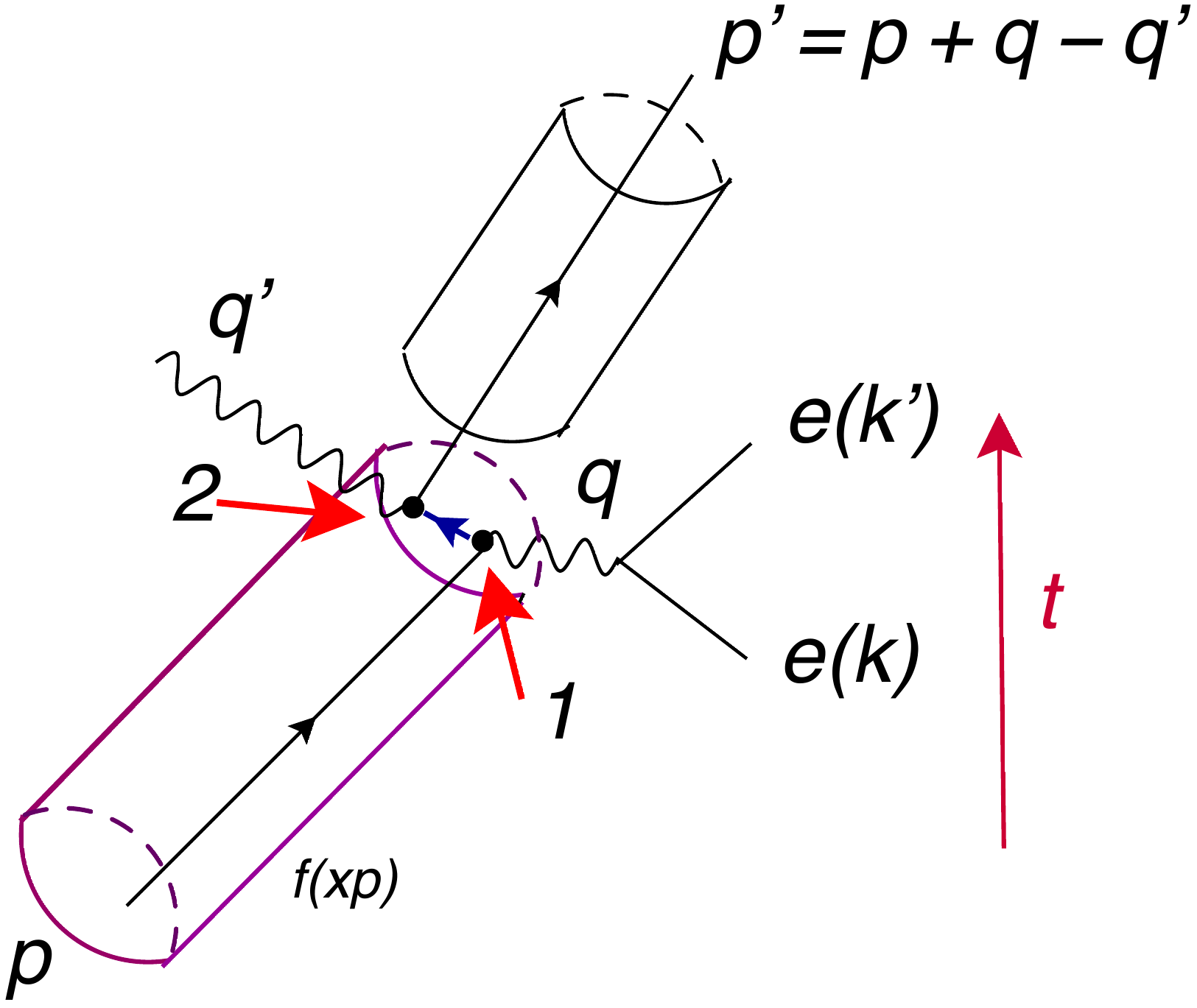}
\caption{The process encoded in a generalized parton distribution.}
\label{fig:gpdf}
\end{figure}

As broad as the forms of Eq.\ (\ref{eq:fact-scheme}) are for inclusive and semi-inclusive cross sections, these are only one route from experiment to the QCD matrix elements that reveal nucleon structure. 
With the matrix element interpretation of parton distributions as inspiration, we can find similar factorized expressions for exclusive amplitudes.   The classic  example is the process of deeply-virtual Compton scattering, illustrated by Fig.\ \ref{fig:gpdf}.   Here we can study a unique quantum-mechanical amplitude, corresponding to the process in the picture.
In this reaction, a proton and an electron collide, and the electron exchanges an off-shell photon with a quark.   The quark then travels for some distance
in the proton, after which it emits an on shell photon.   The proton then reforms itself, with the quark starting out in a new place and with a slightly different momentum.
This prccess is described by matrix elements of the general form, \cite{Ji:1996nm,Belitsky:2005qn,Diehl:2003ny}
\begin{equation}
M \left (x,z,\Delta^2 \right )\ =\ P\cdot n\,
\int \frac{d\lambda}{2\pi}\, e^{i\lambda x P\cdot n}
\langle P+ q |\, \bar Q (- \lambda n)\, \Gamma\, Q(\lambda n)\,  | P \rangle \, ,
\label{eq:gpdf-me}
\end{equation} 
for a quark field $Q$,
in which a relatively small net momentum transfer, $q$ redirects the proton.   This matrix element, corresponding to the figure,
depends on the longitudinal momentum $zp\cdot n$ that the quark loses in the process, along with a corresponding
transverse momentum transfer $\bf \Delta$.   It is from matrix elements such as these that the angular momentum
and similar non-local information on hadronic structure can be formulated in terms of partonic degrees of freedom.

Taken together with the scattered electron, this process corresponds experimentally to
a very distinctive final state, with a slightly deflected nucleon accompanied by an on-shell photon.   The new sets
of nucleon correlations made available in these experiments are beginning to reveal the
stationary nucleon state in the language of partonic degrees of freedom.
Such quantities are also increasingly accessible to lattice calculations \cite{Lin:2017snn,Ebert:2020gxr}.   These are among the most exciting contemporary
confrontations of theoretical with experimental physics of the strong interactions.

\subsection{From jet cross sections to long-time behavior}

The experimental studies described above all depend on grouping together events with similar final states.   For the total DIS cross section, these are {\it all} states that share a scattered electron with the same momentum transfer and energy loss.    For studies of generalized parton distributions, these are the much smaller, more exclusive class of final states with a scattered electron, a high energy photon and a proton.   Another window into short-distance phenomena is provided by sets of states with similar global flows of energy.   These states, characterized by collimated sets of high energy particles, are called jet states \cite{Salam:2010nqg,Dasgupta:2013ihk}. Jet final states also offer a direct picture of processes that take place at the elementary level, the scattering of quarks, the creation of quark pairs, the decay of heavy particles, and the radiation of gluons.

The analysis of jet events has taken on a new life, as we recognize that they should be regarded as a source of` `big data", and hence as a target for ``deep learning" \cite{Datta:2017lxt,Nachman:2021yvi}.   Looking at a detector display, our natural intellegence readily identifies the jets qualitatively, relying on charged particle tracks and such symbolic translations of energy scale as to the length or color of bars representing particles or calorimeter signals.   The information in ensembles of jets, each containing perhaps dozens of particles of various energies and flavor content, is the outcome of the complex process we understand as perturbative showering with nonperturbative corrections, culminating in hadronization \cite{Dobbs:2004qw,Sjostrand:1990we,Forshaw:2020wrq,Dasgupta:2020fwr,Nagy:2020rmk}.   For semi-inclusive or inclusive but ``weighted" cross sections, resummed perturbation theory, often in the language of effective theories, \cite{Becher:2016mmh,Hinderer:2018nkb,Liu:2017pbb,Chien:2019osu}  supplemented by power corrections  \cite{Kang:2018qra,Hoang:2019ceu} can describe a wealth of data.

The flow of energy itself has a field theoretic analog, and much current analysis revolves around exploring the properties of operators involving the energy-momentum tensor \cite{Sveshnikov:1995vi,Korchemsky:1997sy,Korchemsky:1999kt},
\begin{eqnarray}
{\cal E}\left( \hat{n} \right)\ =\
\lim_{R \rightarrow \infty} R^2\, \int_0^\infty dt\, \hat{n}_i\, T_{0i}\left( t,R\hat{n} \right)\, ,
\label{eq:calE_def}
\end{eqnarray}
which measures energy flowing in the specific direction $\hat n$, as seen by a detector.
Correlations between particle jets can be generated from expectations of products of these operators in ``initial" states $|I\rangle$, such as
\begin{equation}
C \left ( E(\hat n_1) \dots E(\hat n_a) , x\right ) \ =\ \langle I | j_\mu(x) \prod_{i=1}^a  {\cal E}\left (\hat{n}_i)\right) j^\mu(0) | I\rangle\, ,
\end{equation}
with $j^\mu$ an electroweak current.   
Such correlations can be calculated perturbatively.  Integrating over angular regions, we can measure energy correlations over differing angular separations \cite{Lee:2019lge}.   

The dependence of angular correlations in energy flow on the medium traversed by partons on their way from a hard scattering to the final state has
been, and remains, a sensitive probe of the properties of hot and cold nuclear media \cite{Arratia:2019vju,Caucal:2020xad}.    Varying energies, reactions, flavors and luminosities will be necessary
to explore fully these fundamental properties of matter.
Interest is also increasing on ``hybrid" expectation values, involving both the energy-momentum tensor and other conserved charges  \cite{Chen:2020vvp,Kang:2020fka}.   The exploration of these
correlations will hopefully lead to new insights on the final stages of the translation from the language of partons into that of hadrons.

\section{Putting the Pieces Together}

What might we expect from theory in the coming years?   Here is a personal perspective, very much ready to be supplemented by novel ideas.

We understand QCD best at its endpoints.   Its dynamics at the very moment of a hard scattering is well-described in perturbation theory, which we can use to evolve to length scales approaching $1/\Lambda_{\rm QCD}$.  At the longest scales, the quark model and effective theories based on its symmetries also afford quantitative descriptions.   In between, perhaps we can think of a process by which current quarks generate mass from radiative energy, becoming constituent quarks.
How to quantify this transition?
Are there observables sensitive to this period?  Photons radiated in this stage might be an example \cite{Wong:2014ila,Kharzeev:2014xta}.

Final states provide a countable amount of information,
and we should be ready to count it all, to the extent possible.
The detailed census within each jet encodes a set of stories.
Jet substructure analyses are beginning the process of breaking this code, but it will surely take many new insights, ideas, and computing power \cite{Larkoski:2017jix,Fraser:2018ieu,Komiske:2020qhg}.

A judicious use of machine learning, new ideas of event display and perhaps quantum information will lead to the
demystification of the transition between the language of partons and that of hadrons.
For amplitudes and  inclusive cross sections, we'll continuously reevaluate to what orders can we calculate.
Analytic fixed-order progress will continue, hand-in-hand with mathematical insight \cite{DelDuca:2011ae,Anastasiou:2016cez,Henn:2019gkr,Caron-Huot:2020grv,Ebert:2020sfi}, while numerically, higher orders will become more accessible for cross sections, and experimental cuts easier to implement.

We'll hopefully see algorithmic evaluations of complex QCD amplitudes and cross sections \cite{Capatti:2020xjc}.
Why might this be possible? It's the magic of unitarity \cite{Lee:1964is,Sterman:1978bj,Frye:2018xjj}, the same conservation of probability that enables us to factorize parton distributions in DIS and makes jet cross sections calculable.
We will then be able to avoid infrared divergences before we integrate over phase space and impose experimental cuts.
We will learn to calculate in four dimensions \cite{Anastasiou:2020sdt,TorresBobadilla:2020ekr} and limit perturbation theory to finite times. This may help make room for a new theory of hadronization.

QCD will be more and more embedded in the Standard Model, with flavor issues coming to the fore at all scales, from $g-2$ all the way to future super high-energy colliders.
Also with the help of the lattice, we will learn more of the correlations within hadrons and in exotic states of matter in the lab and in the universe.
Higher twist will be subsumed into a theory of quark-hadron duality \cite{Shifman:2000jv}, building on the kind of analysis that led to QCD sum rules, relating hadronic properties to the operator product expansion.
A solution to the strong CP may arise in the coming decade, perhaps in connection with an observation of dark matter.
Perhaps what we learn about QCD will suggest other scenarios for early stages of the universe and matter under extraordinary conditions.
This would bring us back to the beginning.

\section{Concluding Thoughts}

In many ways, quantum chromodynamics is the exemplary quantum field theory.  It exhibits highly nontrivial phase structure, accessible now in ion collisions,  and perhaps through its collective flow in smaller systems too.  Its classical solutions, the instantons, opened new perspectives in geometry, and its perturbative amplitudes have echos in the mathematics of  complex analysis and both classical and quantum general relativity.
It serves as a benchmark for more manageable theories, especially conformal quantum field theories and related bootstrap programs.

Even more generally, any answer to the question ``what is quantum field theory?" must encompass and find inspiration from quantum chromodynamics. The quantum field theory of our day is a bit like the calculus of Newton and Leibnitz, not yet fully defined, but extraordinarily effective.   This is one reason why QCD continues to be guided by experiment. For example, only through experience could we know that the particle jets of perturbative QCD
would be there at high energy in the laboratory.

Quantum chromodynamics is approaching its fiftieth year, but for a fundamental theory, it is still young. Newtonian gravity, for example, is still going strong at 334 since it appeared in the {\it Principia}.
Quantum chromodynamics itself remains an astounding discovery, with its mysterious unbroken color symmetry.
Even more, QCD makes it possible for us to face the challenge of bridging the fundamental and the emergent phenomena of the natural world.
This is a hallmark of 21st Century science, and indeed of human thought.

Finally, at the time of this writing, we mark with sadness the loss of Steven Weinberg, whose work and thought echo through these short pages and far beyond, guiding how we think of quantum chromodynamics and so much else in contemporary theoretical physics.   His was a paragon of a life in science.

\section*{Acknowledgements}
Congratulations to the organizers of
DIS XXVIII: 2020 / 2021,
for making it happen, and to all who took part,
with the hope and expectation that DIS XXIX will reflect a regained,
and perhaps reimagined, freedom.

\paragraph{Funding information}
This work was supported in part by the National Science Foundation, through award PHY-1915903.



\bibliography{SciPost_Example_BiBTeX_File.bib}

\nolinenumbers

\end{document}